\documentclass[%
12pt]{iopart}

\usepackage{graphicx}
\usepackage{dcolumn}
\usepackage{bm}
\usepackage{hyperref}
\usepackage[mathlines]{lineno}
\usepackage{float}
\usepackage{subcaption}
\usepackage[font=small,labelfont=bf,justification=justified]{caption}
\usepackage{iopams}
\usepackage{pifont}
\usepackage{amsfonts}

\usepackage{ragged2e}
\usepackage{physics}
\usepackage{enumerate}
\usepackage[usenames,dvipsnames]{color}
\usepackage[shortlabels]{enumitem}
\hypersetup{breaklinks=true}

\begin{document}

\title[Embedding of a non-Hermitian Hamiltonian]{Embedding of a non-Hermitian Hamiltonian to emulate the von Neumann measurement scheme}

\author{Gurpahul Singh$^{1,2,3,*}$, Ritesh K. Singh$^3$ and Soumitro Banerjee$^3$}
\address{$^1$ Perimeter Institute for Theoretical Physics, Waterloo, Ontario N2L 2Y5, Canada}
\address{$^2$ Department of Physics, University of Waterloo, Waterloo, Ontario N2L 3G1, Canada}
\address{$^3$ Department of Physical Sciences, Indian Institute of Science Education and Research, Kolkata 741 246, India }
\ead{singh.pahulg99@gmail.com}
\vspace{10pt}
\begin{indented}
\item[]December 2023
\end{indented}


\begin{abstract}
The problem of how measurement in quantum mechanics takes place has existed since its formulation. Von Neumann proposed a scheme where he treated measurement as a two-part process --- a unitary evolution in the full system-ancilla space and then a projection onto one of the pointer states of the ancilla (representing the ``collapse" of the wavefunction). The Lindblad master equation, which has been extensively used to explain dissipative quantum phenomena in the presence of an environment, can effectively describe the first part of the von Neumann measurement scheme when the jump operators in the master equation are Hermitian. We have proposed a non-Hermitian Hamiltonian formalism to emulate the first part of the von Neumann measurement scheme. We have used the embedding protocol to dilate a non-Hermitian Hamiltonian that governs the dynamics in the system subspace into a higher-dimensional Hermitian Hamiltonian that evolves the full space unitarily. We have obtained the various constraints and the required dimensionality of the ancilla Hilbert space in order to achieve the required embedding. Using this particular embedding and a specific projection operator, one obtains non-Hermitian dynamics in the system subspace that closely follow the Lindblad master equation. This work lends a new perspective to the measurement problem by employing non-Hermitian Hamiltonians.
\end{abstract}

\vspace{2pc}



\section{Introduction}
Since the very beginning, the measurement problem in quantum mechanics has troubled many physicists and philosophers. A lot of models and interpretations have been proposed trying to explain the process of measurement and the collapse of the wavefunction in particular. The Copenhagen interpretation \cite{copenhagen} was one of the first among these attempts that put together the ideas of Niels Bohr, Werner Heisenberg, Max Born and others. It states that a quantum state is intrinsically indeterminate -- a superposition of possible eigenstates corresponding to an observable-- until a measurement is made, and the act of measurement \textit{collapses} it to one of those eigenstates. Some of the models that came later were the De-Broglie Bohm theory \cite{BohmQM}, Everett many-worlds interpretation\cite{EverettManyworld}, von Neumann-Wigner interpretation \cite{Wigner'sfriend}, Transactional interpretation\cite{Transactional_interpretation}, Ghirardi-Rimini-Weber (GRW) theory \cite{GRWmodel}, Continuous Spontaneous Localization (CSL) model \cite{CSLmodel1, CSLmodel2}, Diosi-Penrose (DP) model \cite{Diosi, Penrose1996}, Relational QM \cite{RelationalQM} and decoherence model of collapse \cite{Decoherence}. 

Eventually, the Copenhagen interpretation \cite{copenhagen} prevailed. Following it, von Neumann proposed a measurement scheme \cite{nielsen_chuang_2010, sakurai_napolitano_2017}. The initial system-ancilla state is a product state comprising the system state and the pointer states that make up the ancilla. This state becomes entangled during the measurement process due to a unitary operation, and then a projection operator represents the collapse of the system subspace onto one of the eigenstates of the measurement operator. 

On the other hand, open quantum systems are well described by a differential equation in the (system) density matrix formalism, obtained after partial tracing over the environment degrees of freedom. This Lindblad master equation \cite{Lindblad1, Lindblad2, Lindblad3, Lindbladsummary} brings about decoherence in the system subspace such that the final state is a thermal one. In the context of measurement, if the jump operators are Hermitian, we get a probability distribution of being in the eigenstates of the measurement operator \cite{Hermitianladdder}. This is similar to the first part of the measurement scheme proposed by von Neumann---  unitary evolution in the full space while non-unitary evolution in the system subspace.

We follow the von Neumann scheme and propose that the measurement process occurs over a finite interval. During this interval, the apparatus interacts with the system, and the whole dynamics is governed by a Hamiltonian acting on the combined Hilbert space of the system and the apparatus. This leads to \textit{leaking} of probabilities from the system to the apparatus and vice-versa. When seen from the point of view of the system Hilbert space alone, the dynamics appears to be governed by a non-Hermitian Hamiltonian.

Non-Hermitian quantum mechanics has been an extensive topic of research, both theoretically and experimentally, in recent years \cite{PTreview1, PTreview2}. In the context of measurement, physicists have tried to explore the connection between weak measurement and non-Hermitian operators \cite{WeakmeasurementnonHermitian, WeakmeasurementnonHemitian2, WeakmeasurementnonHemitian3}. There have also been attempts to describe open quantum systems effectively in terms of non-Hermitian Hamiltonians \cite{NonHermitianappliedtoopen1, NonHermitianappliedtoopen2, NonHermitianappliedtoopen3, NonHermitianappliedtoopen4, NonHermitianappliedtoopen5, NonlinearvonNeumann, NonHermitianappliedtoopen6, NonHermitianappliedtoopen7, NonHermitianappliedtoopen8}. Specifically, the link between the Lindblad master equation and non-Hermitian dynamics has been studied before \cite{LindbladandnonHermitian, LindbladandnonHermitian2, Lindbladconn}. 

For our model, we propose that the system subspace dynamics is governed by a non-Hermitian Hamiltonian (non-unitary evolution) while the full system-ancilla space dynamics is dictated by a Hermitian Hamiltonian (unitary evolution). For this, we use the Naimark dilation protocol \cite{Dilationbook} to dilate a non-Hermitian Hamiltonian into a higher-dimensional Hermitian one. This protocol, also called the embedding formalism, has already been used previously to embed $\mathcal{PT}$-symmetric non-Hermitian Hamiltonians into Hermitian Hamiltonians \cite{Sciencepaper, NaimarkDilation1, NaimarkDilation2, Embedding1, Embedding2, QMBT1SDas}. Using this, we try to emulate the dynamics obtained in the first part of the von Neumann measurement scheme. We find the various constraints on the non-Hermitian Hamiltonian itself in order for it to be embedded properly. To illustrate our model, we consider a two-level system state coupled with a two-level ancilla. After obtaining the 4-dimensional Hermitian Hamiltonian using the embedding formalism, the final state after the measurement converges to one of the eigenvectors with a certain probability but not the other. We increase the dimensionality of the ancilla Hilbert space to 4 and thereafter repeat the calculations to get as close to mimicking the measurement results as possible. The formalism has also been generalized for $N$-level system state. The entropy, eigenstate probabilities, and Bloch sphere trajectories are studied accordingly in order to compare the obtained dynamics with the Lindblad evolution.

The organization of the paper is as follows: in Sec.~\ref{sec:2}, we give a brief outline of the von Neumann scheme of measurement, the Lindblad master equation, and non-Hermitian dynamics. In Sec.~\ref{sec:3}, we review the embedding protocol and obtain the constraints that must be satisfied for its validity. We also discuss how the choice of the system Hamiltonian (non-Hermitian) is affected due to the constraint. In Sec.~\ref{sec:4}, we demonstrate our protocol for a two-dimensional system Hamiltonian coupled with a two-level ancilla and plot the relevant graphs. In Sec.~\ref{sec:5}, we extend our analysis by introducing a four-level ancilla such that the full space is now eight-dimensional and then generalize the same to $N$-dimensions. In Sec.~\ref{sec:6} and \ref{sec:7}, we discuss the efficiency and accuracy of our model along with the associated challenges and summarize the results of our work. 

\section{The von Neumann scheme}\label{sec:2}
There are a lot of interpretations and mechanisms for how a measurement takes place. But one of the most popular approaches is one given by von Neumann \cite{nielsen_chuang_2010}.

Let $\{M_k\}$ be the measurement operators on the state space of the system. The system state is given by $|\psi\rangle$ which lives in the Hilbert space $\mathcal{H}_s$ while the full state is given by $|\Psi\rangle$ which lives in the system-ancilla Hilbert space $\mathcal{H}_{sa}$. Let the ``pointer states” of the apparatus initially be an arbitrary collection of all possible outcomes that the apparatus can finally point to.

Then the initial full state is a product state between the system and apparatus states
\begin{equation}\label{eq:01}
    |\Psi\rangle = |\psi\rangle\otimes(|0\rangle+\dots|n\rangle)_a
\end{equation}

During the measurement, this full state is acted upon by a unitary operator $\mathcal{U}$ such that the system and ancilla become entangled.
\begin{equation}\label{eq:02}
    \mathcal{U}|\Psi\rangle = |\Psi'\rangle = \sum_{k=0}^n M_k|\psi\rangle\otimes|k\rangle_a
\end{equation}

The final state $|\Psi'\rangle$ is an entangled state which gives a superposition of all possibilities a state could be in.

For example, let us have $M_{0,1} = P_{0,1}$, where $P_0 = |0\rangle\langle 0|$ and $P_1 = |1\rangle\langle 1|$  are the projection operators to $|0\rangle$ and $|1\rangle$ state. Suppose the initial state is $|\psi(0)\rangle = c_1|0\rangle +c_2|1\rangle$. Then, according to the above scheme,  
\begin{equation}\label{eq:03}
    |\Psi'\rangle = c_1|0\rangle_s|0\rangle_a + c_2|1\rangle_s|1\rangle_a
\end{equation}
which reads as follows: after the measurement, the final state is either $|0\rangle$ and the apparatus also points to $|0\rangle$ (with probability $|c_1|^2$) or the final state is $|1\rangle$, and the apparatus also reads $|1\rangle$ (with probability $|c_2|^2)$. 

The implication of this is that if we have an ensemble of initial states and we do a measurement, we will get $|c_1|^2$ particles going to $|0\rangle$ state and $|c_2|^2$ particles going to $|1\rangle$ state. In the density matrix formalism, we would obtain the final state density matrix, which is diagonal with $|c_1|^2$ and $|c_2|^2$ as the diagonal elements.

For a single particle experiment, it can be either one way or the other. Von Neumann says that there is a projection operator 
$$\mathcal{P}_m = \mathbb{I}\otimes|m\rangle_a\langle m|_a$$ 
which acts on the full state so that the final state obtained is $M_m|\psi\rangle$ up to normalization. The post-measurement system state would look like
$$\frac{M_m|\psi\rangle}{\sqrt{\langle\psi|M_m^{\dagger}M_m|\psi\rangle}}$$
Say, if $\mathcal{P}_0 = \mathbb{I}\otimes|0\rangle_a\langle 0|_a$, then the post-measurement state in the above example \eqref{eq:03} would have been
$$\mathcal{P}_0|\Psi'\rangle\longrightarrow e^{i\theta}|0\rangle\xrightarrow{\mathcal{P}_0} e^{i\theta}|0\rangle $$
where $e^{i\theta} = \frac{c_1}{|c_1|}$. Upon applying the $\mathcal{P}_0$ operation the second time, we get back to the same state. And this is what is termed the ``collapse" of the wavefunction.

\subsection{Lindblad Master Equation}
The Lindblad master equation tends to explain the evolution from the product state $|\Psi\rangle$ to the entangled state $|\Psi'\rangle$ using a first-order linear differential equation of the system density matrix $\rho_s$. The equation looks like
\begin{equation}\label{eq:04}
    \dot\rho_s = -i[H_s,\rho_s] +\sum_k\Gamma_k\left(L_k\rho_s\L^{\dagger}_k-\frac{1}{2}\{ L^{\dagger}_k L_k,\rho_s\} \right)
\end{equation}
where $H_s$ is the Hamiltonian for the system state $\rho_s$, $L_k$ are called jump operators, which come in due to the influence of the environment and thermalize the state. The factor $\Gamma_k$ determines how fast the thermalization happens. 

Now, for example, to measure spin in the $\hat{n}$ direction, we use $L_k = \sigma_n$ and evolve an initial state using the above equation. The final density matrix yields the probabilities of being in $\ket{\uparrow_n}$ and $\ket{\downarrow_n}$ as the diagonal elements (decoherence) \cite{Hermitianladdder}. Thus, the Lindblad evolution brings about the non-unitary evolution of a system state while the full system-apparatus state undergoes unitary evolution under some operator $\mathcal{U}$. 

We would like to see if there is a Hamiltonian formalism to bringing about decoherence and obtaining an entangled final state during measurement. We know that non-Hermitian Hamiltonians evolve states in a non-unitary fashion. It would be worth investigating if, by using a non-Hermitian Hamiltonian $H_s$ in the system subspace and then embedding it into a Hermitian Hamiltonian $H_{sa}$ that governs the full state, we can obtain an entangled final state $|\Psi'\rangle$ that is given by the von Neumann scheme Eq.~\eqref{eq:02}. Then, in some sense, we would have emulated the decoherence effect obtained from the Lindblad master equation but in a Hamiltonian formalism. The $\mathcal{U}$ operator, in that case, would become  
\begin{equation}
    \mathcal{U} = e^{-i\int H_{sa}dt}
\end{equation}

In order to obtain these Hamiltonians, let us look at how states evolve when governed by a non-Hermitian Hamiltonian. 

\subsection{Non-Hermitian Hamiltonians}\label{III}
Let us consider a non-Hermitian Hamiltonian $H$ of the form $H = H_h - iH_a$ where $H_h$ and $H_a$ are the Hermitian and anti-Hermitian parts. Also, $H_h^{\dagger} = H_h$ and $H_a^{\dagger} = H_a$. The density matrix $\rho$ of a system, evolved via the Hamiltonian $H$ will follow \cite{NonlinearvonNeumann}
\begin{equation}\label{eq:06}
    \dot\rho = -i[H_h,\rho]-\{H_a,\rho\} + 2\:\text{tr}(\rho H_a)\rho
\end{equation}
which is similar to the nonlinear Schr\"odinger equation mentioned in \cite{Gisin}. The nonlinearity comes in because of the last term on the RHS introduced to preserve the trace of $\rho$. 
The evolved state $\rho(t)$ is then given by
\begin{equation}\label{eq:07}
    \rho(t) = \frac{e^{-iHt}\rho(0)e^{iH^{\dagger}t}}{\text{Tr}(e^{-iHt}\rho(0)e^{iH^{\dagger}t})}
\end{equation}
Several properties of Eq.~\ref{eq:06} have been listed in \cite{NonlinearvonNeumann}. The one that would be of importance to us is that if $H$ has complex eigenvalues, then the eigenvector with the largest imaginary part of the eigenvalue becomes the attractor (sink) of the dynamics \cite{Ownpaper}. This means that any initial state evolved via $H$ will eventually converge to this eigenvector. 

The expression of a time-evolved density matrix $\rho(t)$ when evolved using a diagonal non-Hermitian Hamiltonian $H$ has been shown in \cite{Ownpaper}. The significance of choosing a diagonal Hamiltonian will be clear in the later sections. For now, diagonalizing a Hamiltonian has no effect on the nature of evolution that it dictates.

Let us take the initial state to be 
$$\rho(0) = \left [
 \begin{array}{cc}
 |c_1|^2&c_1c_2^* \\
 c_1^*c_2&|c_2|^2 \\ 
 \end{array}
 \right ]$$
 Broadly, there can be three ways to send an initial state $\rho$ to one of the eigenvectors of a diagonal Hamiltonian $H$, i.e., to $|0\rangle$ or $\ket{1}$. All the 3 cases below deal with Hamiltonians that send an initial state to $|0\rangle$. 


\textbf{Case A:} We add an imaginary number $i\gamma$ ($\gamma>0$) to one eigenvalue and subtract it from the other. This is like ``pushing" the state towards one eigenvector and, at the same time ``pulling" it away from the other. So, $H$ would look like
\begin{align}
    \label{eq:09}
        H^A &= \text{diag}(\lambda_1+i\gamma,\lambda_2-i\gamma)
\end{align}
When time evolved using Eq.~\eqref{eq:07}, the density matrix looks like
\begin{equation}
 \rho^A(t) \!=\!\! \left [ \!
 \begin{array}{cc}
 \dfrac{|c_1|^2}{|c_1|^2\!+\!|c_2|^2e^{-4\gamma t}}& \dfrac{c_1c_2^*e^{-i2\omega t}}{|c_1|^2e^{2\gamma t}\!+\!|c_2|^2e^{-2\gamma t}} \\
 \dfrac{c_1^*c_2e^{i2\omega t}}{|c_1|^2e^{2\gamma t}\!+\!|c_2|^2e^{-2\gamma t}} & \dfrac{|c_2|^2}{|c_1|^2e^{4\gamma t}\!+\!|c_2|^2}. \\ 
 \end{array}
 \!\! \right ]
\end{equation}
It is clear from the above equation that as $t\rightarrow \infty$, the off-diagonal terms go to 0 (decoherence). The $\rho^A_{22}(t)$ element also goes to 0 while the $\rho^A_{11}(t)$ term goes to 1. Hence, the state ultimately approaches $|0\rangle$ state given enough time. 

\textbf{Case B:} We add $i\gamma$ to the upper diagonal element, which is like ``pushing" the state towards $|0\rangle$ without pulling it from $|1\rangle$. 
$H^B$ looks like
$$H^B = \text{diag}(\lambda_1+i\gamma,\lambda_2)$$

In this case, $\rho^B(t)$ turns out to be exactly same as $\rho^A(t)$ except that $\gamma$ is replaced by $\gamma/2$. So, the rate of convergence is halved.

\textbf{Case 
 C:} Let us subtract $-i\gamma$ from the lower diagonal element. In this case
$$H^C = \text{diag}(\lambda_1,\lambda_2-i\gamma)$$
For this case, $\rho^{C}(t) = \rho^B(t)$, which means there is no difference in just pushing a state to an eigenvector or just pulling it away from the other.

\textbf{Case D:} In this case, the Hamiltonian used is $H^C$, but we would not use the trace normalization in the evolution equation \eqref{eq:07}. The density matrix expression now looks like 
\begin{equation}\label{eq12}
     \rho^{D}(t) = \left [ 
 \begin{array}{cc}
 |c_1|^2& e^{-\gamma t}c_1c_2^*e^{-i2\omega t} \\
 e^{-\gamma t}c_1^*c_2e^{i2\omega t}& |c_2|^2e^{-2\gamma t} \\ 
 \end{array}
  \right ]
\end{equation}

Here, we see that all the terms go to 0 as $t\rightarrow \infty$ except $\rho^{D}_{11}(t)$ term, which stays at $|c_1|^2$. Therefore, if the density matrix represented an ensemble of particles, this evolution would leave the population in the ground state intact. There would be decay in the excited state population and also decoherence. 

As mentioned before, the Lindblad master equation causes decoherence such that, given enough time, the density matrix state thermalizes to a mixed state with no off-diagonal terms. The diagonal terms sum up to one, and each of those gives the population at different energy levels.   

Case D leaves out a density matrix whose trace is not preserved, which means the particle number is not preserved. The ground state population remains as it is. But if we use a different Hamiltonian with $-i\gamma$ in the upper diagonal, we would be left with just the excited state population in the density matrix. Thus, the sum of these two dynamics has a unit trace and looks like the density matrix obtained from Lindblad-type dynamics. 

This means that the Lindblad Markovian dynamics can be obtained as an incoherent sum of two different non-Hermitian Hamiltonian dynamics. 

We have obtained a non-Hermitian Hamiltonian that leads to decoherence and sends an initial state to one of the eigenvectors of the Hamiltonian with the respective probability. But for an $N$-level system state, there must be $N$ number of Hamiltonians that access all of the eigenvectors of the Hamiltonian. We would try and see if there is a way to embed these $N$ Hamiltonians into a single Hermitian Hamiltonian that operates on the full system-ancilla state $|\Psi\rangle$ following the von Neumann scheme mentioned in the previous section. For that, we will first discuss the embedding protocol in the next section.

\section{The Embedding protocol}\label{sec:3}
The dilation of a general $N$-dimensional non-Hermitian Hamiltonian into a $2N$-dimensional Hermitian Hamiltonian has been done in \cite{Sciencepaper} with details mentioned in the supplementary texts. We have found certain conditions on the lower dimensional Hamiltonian so that the higher dimensional Hamiltonian remains Hermitian for all times. Before going into that, we briefly review the embedding protocol mentioned in \cite{Sciencepaper}.

The dynamics of the system state $|\psi\rangle$ is governed by the non-Hermitian system Hamiltonian $H_s(t)$, which can be time-dependent or time-independent. 
\begin{equation}\label{eq:1}
    i\frac{d}{dt}|\psi(t)\rangle = H_s(t)|\psi(t)\rangle
\end{equation}
The dilated system-ancilla state is taken as 
\begin{equation}\label{eq:2}
    |\Psi(t)\rangle = |\psi(t)\rangle \ket{\downarrow_n} + \eta(t)|\psi(t)\rangle\ket{\uparrow_n}
\end{equation}
where $\eta(t)$ is a linear operator and $\ket{\uparrow_n}$, $\ket{\downarrow_n}$ are the eigenstates of a general Pauli operator $\sigma_{\hat{n}} = \hat{n}\cdot\vec{\sigma}$, which represent the ancilla states. This state $|\Psi(t)\rangle$ is driven by the dilated time dependent Hermitian Hamiltonian $H_{sa}(t)$ : 
\begin{equation}\label{eq:3}
    i\frac{d}{dt}|\Psi(t)\rangle = H_{sa}(t)|\Psi(t)\rangle
\end{equation}
We can expand $H_{sa}(t)$ as follows:
\begin{align}
    \label{eq:4}
    \begin{split}
         H_{sa}(t) = H_{sa}^{(\uparrow\uparrow)}(t)\otimes\ket{\uparrow_n}\bra{\uparrow_n} + H_{sa}^{(\uparrow\downarrow)}(t)\otimes\ket{\uparrow_n}\bra{\downarrow_n}
         \\
   +H_{sa}^{(\downarrow\uparrow)}(t)\otimes\ket{\downarrow_n}\bra{\uparrow_n}+H_{sa}^{(\downarrow\downarrow)}(t)\otimes\ket{\downarrow_n}\bra{\downarrow_n}
    \end{split}
\end{align} 
For $H_{sa}(t)$ to be Hermitian, we require:
$$(H_{sa}^{(\uparrow\uparrow)}(t))^{\dagger} = H_{sa}^{(\uparrow\uparrow)}(t) \hspace{5pt};\hspace{5pt} (H_{sa}^{(\downarrow\downarrow)}(t))^{\dagger} = H_{sa}^{(\downarrow\downarrow)}(t)$$
$$(H_{sa}^{(\uparrow\downarrow)}(t))^{\dagger} = H_{sa}^{(\downarrow\uparrow)}(t)$$
Demanding this Hermiticity, the evolution for $\eta(t)$ is obtained as:
\begin{equation}\label{eq:7}
    i\frac{d}{dt}[\eta^{\dagger}(t)\eta(t)] = H_s^{\dagger}(t)[\eta^{\dagger}(t)\eta(t)+\mathbb{I}]-[\eta^{\dagger}(t)\eta(t)+\mathbb{I}]H_s
\end{equation}
Defining $M(t) \equiv \eta^{\dagger}(t)\eta(t) +\mathbb{I}$, one gets
\begin{equation}\label{eq:8}
    i\frac{d}{dt}M(t) = H_s^{\dagger}(t)M(t)-M(t)H_s(t)
\end{equation}
So, $\eta(t) = U(t)(M(t) - \mathbb{I})^{\frac{1}{2}}$ for some unitary matrix $U(t)$.  $M(t)$ is Hermitian (by construction), and its evolution can be written as 
\begin{equation}\label{eq:9}
    M(t) = \mathcal{T}e^{-i\int H_s^{\dagger}(t)dt}M(0) \Bar{\mathcal{T}}e^{i\int H_s(t)dt}
\end{equation}
where $\mathcal{T}$ and $\Bar{\mathcal{T}}$ are time-ordering and anti-time-ordering operators, respectively. Also, the definition of $M(t)$ puts a constraint on it that its lowest eigenvalue should remain more than 1 throughout the evolution.

\subsection{A particular choice}
Let us take $\hat{n} = \hat{y}$ such that ancilla states are eigenstates of $\sigma_y$ with $\ket{\downarrow_y} = |-\rangle_y$ and $\ket{\uparrow_y} = |+\rangle_y$. We have some degree of freedom with regard to the choice of $U(t)$ and $H_{sa}^{(++)}(t)$. In \cite{Sciencepaper}, the authors have made two specific choices. (i) $U(t) = \mathbb{I}$, which means that $\eta(t)$ is Hermitian and commutes with $M(t)$. 

(ii) The second choice is made on $H_{sa}^{(++)}(t)$ for easy construction in an NV center.
\begin{equation}\label{eq:10}
    H_{sa}^{(++)}(t) = \{H_s(t)+[i\frac{d}{dt}\eta(t)+\eta(t)H_s(t)]\eta(t)\}M^{-1}(t)
\end{equation}
It can be checked this choice of $H_{sa}^{(++)}(t)$ is Hermitian. From these two choices, one obtains $H_{sa}^{(--)}(t) = H_{sa}^{(++)}(t)$ and $H_{sa}^{(+-)}(t) = -H_{sa}^{(-+)}(t)$. Thus, collecting the terms, we arrive at a simplified expression for $H_{sa}(t)$:
\begin{equation}\label{eq:11}
    H_{sa}(t) = A(t)\otimes\mathbb{I} + B(t)\otimes\sigma_z 
\end{equation}
where 
\begin{equation}\label{eq:12}
    A(t) = \{H_s(t)+[i\frac{d}{dt}\eta(t)+\eta(t)H_s(t)]\eta(t)\}M^{-1}(t)
\end{equation}
\begin{equation}\label{eq:13}
    B(t)= i[H_s(t)\eta(t)-\eta(t)H_s(t)-i\frac{d}{dt}\eta(t)]M^{-1}(t)
\end{equation}

We consider some $H_s(t)$ and choose $M(0) = m_0\mathbb{I}$ to begin with, where $m_0$ is some positive number. Time evolved $M(t)$ is obtained using \eqref{eq:9} and hence $A(t)$ and $B(t)$. After evolving the system-ancilla state $|\Psi(t)\rangle$ via  $H_{sa}(t)$ obtained as above, we can infer the state of the system $|\psi(t)\rangle$ at time t by applying the normalized projection $\mathcal{P}$ on $|\Psi(t)\rangle$ at that instant t.
\begin{equation}\label{eq:14}
    \mathcal{P}|\Psi(t)\rangle = |\psi(t)\rangle\otimes|-\rangle_y = \mathcal{T}e^{-i\int H_s(t)dt}|\psi(0)\rangle\otimes|-\rangle_y
\end{equation}
where 
\begin{equation}\label{eq:proj}
    \mathcal{P} = \dfrac{\mathbb{I}\otimes|-\rangle_y \langle -|_y}{\sqrt{\langle\Psi(t)|(\mathbb{I}\otimes|-\rangle_y\langle -|_y)|\Psi(t)\rangle}}.
\end{equation} 

Post-selecting over the ancilla state $|-\rangle_y$, we get the evolution of state $|\psi(t)\rangle$, which is governed by just the non-Hermitian Hamiltonian. The time-dependent normalization in the projection operator preserves the norm of the state $|\psi(t)\rangle$. 

The non-Hermitian system Hamiltonian $H_s(t)$ needs to be a specific choice in order to ensure that the eigenvalues of $M(t)$ be greater than one during the evolution. This can be seen from equation \eqref{eq:9}. Let us see what conditions we get on a general $N$-dimensional $H_s(t)$ given the constraint of the eigenvalues of $M(t)$. 

\subsection{Eigenvalue constraint on $M(t)$}
Let us, for now, assume the system Hamiltonian to be time-independent $H_s$. We can expand it as follows \cite{Georgi:1982jb}
\begin{equation}
    H_s = r_0\mathbb{I}_N + \sum^{N^2-1}_{j=1} r_jT_j
\end{equation}
where the coefficients $r_0$ and $r_j$ ($j=1,2,\dots,N^2-1$) are complex and $T_j$ are the generators of $\text{SU}(N)$ which are traceless and Hermitian. Out of these $N^2-1$ generators, $N-1$ are diagonal. Let us denote those by $D_j$. So, if we diagonalize $H_s$, we can expand it as follows
$$H_s = d_0\mathbb{I}_N + \sum_{j=1}^{N-1}d_jD_j$$
The coefficients $d_0,d_j$ are complex such that they form each of the diagonal elements of $H_s$. Let us write the diagonal $H_s$ as 
$$H_s = \text{diag}(\lambda_1 +i\gamma_1, \lambda_2+i\gamma_2,\dots, \lambda_N+i\gamma_N)$$
where $\lambda_k$ and $\gamma_k$ represent the real and imaginary parts of the eigenvalues of $H_s$ respectively. Now, since we have the freedom to choose $M(0)$, we assume it commutes with $H_s$ and hence with $H_s^{\dagger}$. So, starting from Eq.~\eqref{eq:9},
$$M(t) = M(0)e^{-i H_s^{\dagger}t}e^{iH_s t}$$
\begin{align}
\begin{split}
    M(t) = M(0)\text{diag}(e^{(-i\lambda_1 -\gamma_1)t},\dots, e^{(-i\lambda_N-\gamma_N)t})\cross\\\text{diag}(e^{(i\lambda_1 -\gamma_1)t},\dots, e^{(i\lambda_N-\gamma_N)t})
\end{split}
\end{align}
$$\Rightarrow M(t) = M(0)\text{diag}(e^{-2\gamma_1t},e^{-2\gamma_2t},\dots, e^{-2\gamma_Nt})$$
Thus, it is clear that in order for the eigenvalues of $M(t)$ to remain greater than one for all times, each of the $\gamma_k$ must be non-positive. Let us see this using a two-level example. 

Let us take $H_s = \sigma_z-i\gamma\frac{(\mathbb{I}+\sigma_z)}{2}$. This choice is exactly the same as Hamiltonian $H^C$ considered in section~\ref{III} where we subtracted $i\gamma$ from the lower diagonal element so that an initial state is sent to $\ket{0}$. The initial $M(0) = m_0\mathbb{I}$ as in \cite{Sciencepaper} where $m_0>1$. 
The expression for eigenvalues of $M(t)$ using Eq.~\ref{eq:9} comes out to be
$$\lambda_{\pm} = m_0e^{\gamma t}(\cosh{(\gamma t)}\pm\sinh{(\gamma t)})$$
$$\Rightarrow \lambda_+ = m_0 e^{2\gamma t}\hspace{1cm}\lambda_- = m_0$$
using $\cosh{(x)}\pm\sinh{(x)} = e^{\pm x}$. Thus, this is a valid $H_s$ for all times since the lowest eigenvalue remains greater than one throughout the evolution. As pointed out, this is the same Hamiltonian that led us to establish a connection with the Lindblad master equation evolution in section~\ref{III}. Interestingly, the same Hamiltonian satisfies the eigenvalue constraints for $M(t)$ for all times. 

\section{Two-level state evolution via diagonal $H_s(t)$}\label{sec:4}
We have seen that a particular diagonal choice of $H_s(t)$ respects the eigenvalue constraint on the lowest eigenvalue of $M(t)$ for all times. Also if $H_s(t)$ is diagonal, then $H_s(t)$, $M(t)$ and $\eta(t)$ commute with each other, so that from \eqref{eq:12} and \eqref{eq:13}, we have simpler expressions for $A(t)$ and $B(t)$,
\begin{equation}\label{eq:19}
    A(t) = \dfrac{H_s^{\dagger}(t)+ H_s(t)}{2}
\end{equation}
\begin{equation}\label{eq:20}
    B(t) = \dfrac{H_s^{\dagger}(t)- H_s(t)}{2i}\eta^{-1}(t)
\end{equation}

To emulate the von Neumann scheme, an initial state $|\psi\rangle$ must be sent to each of the eigenvectors such that the final state is an entangled state as in Eq.~\eqref{eq:02}. For a 2-dimensional Hilbert space, we know that using an appropriate non-Hermitian Hamiltonian $H_s$ (the one shown in case C in section~\ref{III}), one can send an initial state to $|0\rangle$. Also, the same Hamiltonian respects the eigenvalue constraints for the embedding protocol. Hence, we will be using that particular Hamiltonian for our further discussion. But the non-Hermiticity enters after some initial time $t_i$ (just when the measurement starts) and then exits when the measurement ends at time $t_f$. 

To construct what has been described above, we have chosen $H_s(t)$ to be comprised of a Hermitian $H_h$ and a non-Hermitian part $H_{nh}(t)$. The latter is a switching function that switches on at some time $t_i$ and then switches off at $t_f$. For simplicity, the rate of switching on and the amplitude of this part depends on a single parameter $\gamma$.
\begin{equation}\label{eq:21}
    H_{s_{\pm}}(t) = H_h+H_{nh_{\pm}}(t)
\end{equation}
$$H_{nh_{\pm}}(t) = f(t)H_{m_{\pm}}$$
where $f(t)$ is the switching function chosen as
\begin{equation}\label{eq:ft}
    f(t) = -i\frac{\gamma}{2}[\tanh(\gamma(t-t_i))-\tanh(\gamma(t-t_f))]
\end{equation}

\begin{figure*}
    \centering
    \begin{subfigure}[b]{0.34\textwidth}
        \includegraphics[width=\textwidth]{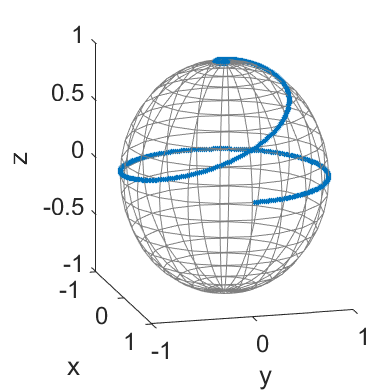}
        \caption{}
        \label{fig2a}
    \end{subfigure}
    \begin{subfigure}[b]{0.44\textwidth}
        \includegraphics[width=\textwidth]{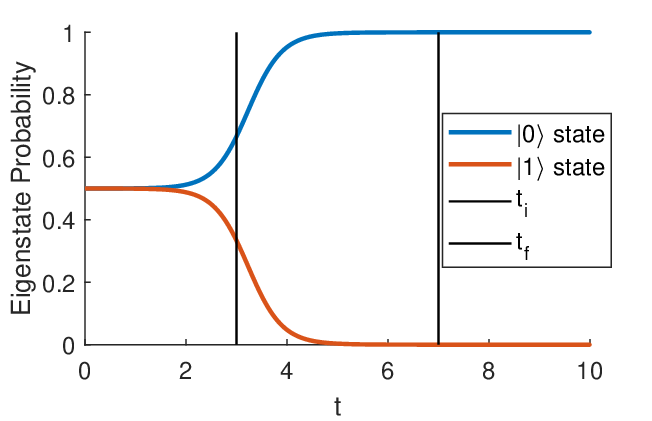}
        \caption{}
        \label{fig2b}
    \end{subfigure}
     \caption{\justifying{(a) The evolution of system state $|\psi(t)\rangle$ as seen on the Bloch sphere. We can see that the initial state ($|+\rangle$ eigenstate) converges to the North pole or $|0\rangle$ state finally. (b) This shows the probability of being in the eigenkets $|0\rangle$ (blue) and $|1\rangle$ (orange) of $H_h$. The black lines mark the switching on $t_i=3$ and switching off $t_f=7$ times for the non-Hermitian part $H_{nh_{-}}(t)$. $M(0) = m_0$, where $m_0 = 1+\epsilon$, $\epsilon$ is a very small positive number.}}
    \label{fig2}
\end{figure*}

We choose both parts to be diagonal with $H_h = \sigma_z$ and $H_{m_{\pm}} = \dfrac{\mathbb{I}\pm\sigma_z}{2}$. The sign in $H_{m_{\pm}}$ dictates whether the system state $|\psi\rangle$ converges to $|0\rangle$ or $|1\rangle$ (North pole or South pole of the Bloch sphere). In FIG.~\ref{fig2a}, it can be seen that the initial state $|\psi(0)\rangle=|+\rangle$ converges to $|0\rangle$ when $H_{m_{-}}$ is applied. If we do the same with $H_{m_{+}}$, it goes to the South pole. 

We can also calculate the probability of converging to the eigenvector $|0\rangle$ or $|1\rangle$ by calculating $|\langle\psi(t)|0\rangle|^2$ or $|\langle\psi(t)|1\rangle|^2$ respectively. These probabilities are shown in FIG.~\ref{fig2b}. We can see that for $H_{m_{-}}$, the state finally goes to $|0\rangle$ eigenket while the opposite will happen for $H_{m_{+}}$.

\begin{figure}
    \centering
    \includegraphics[scale=0.8]{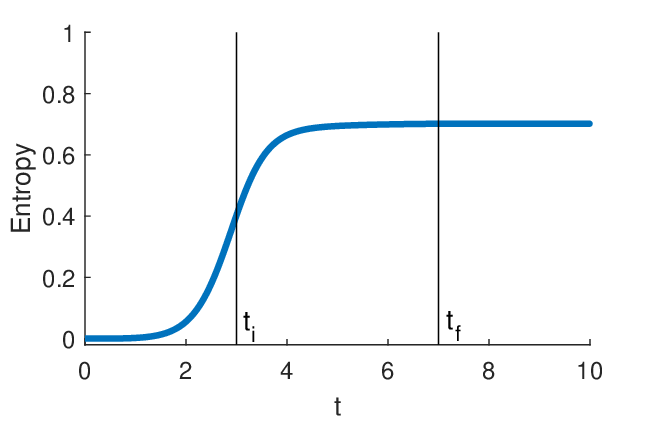}
    \caption{\justifying{The von-Neumann entropy, calculated using equation \eqref{eq:22}, is 0, in the beginning, indicating a product state (since $\eta(0) \approx \mathbb{I}$). The entanglement increases when the non-Hermiticity is switched on and saturates when $|\psi_s(t)\rangle$ has converged to $|0\rangle$.}}
    \label{fig3}
\end{figure}

Suppose $\rho_{sa}(t)$ is the density matrix for the full system-ancilla state. Then, we can find the reduced density matrix for the system $\rho_s(t) = \text{Tr}_a(\rho_{sa}(t))$. The von Neumann entropy, which quantifies the entanglement between system and ancilla states, becomes
\begin{equation}\label{eq:22}
    S = -\text{Tr}(\rho_s(t)\log(\rho_s(t)))
\end{equation}
This has been shown in FIG.~\ref{fig3}. The initial product state becomes entangled when $H_{nh}(t)$ is switched on. The entropy saturates after some time and remains at that value even as $H_{nh}(t)$ is switched off.


The dynamics are similar to a two-level state evolving independently via non-Hermitian Hamiltonian $H_s(t)$. The dynamical equation followed by this two-level state is the nonlinear von Neumann equation \eqref{eq:06}. Because of the time-dependent projection that we have introduced in our embedding protocol \eqref{eq:14}, the projected state dynamics of $\ket{\psi(t)}$ follow exactly the equation \eqref{eq:07}.
 
To calculate the speed of convergence, we will use the definition as in \cite{trifonov2004geometric} where the distance between two arbitrary pure states $\psi_1$ and $\psi_2$ is
\begin{equation}\label{eq:24}
    \delta = \cos^{-1}(|\langle\psi_1|\psi_2\rangle|)
    \Rightarrow \delta = \cos^{-1}(\sqrt{|\text{Tr}(\rho_1\rho_2)|})
\end{equation}

\begin{figure}
    \centering
    \includegraphics[scale=0.8]{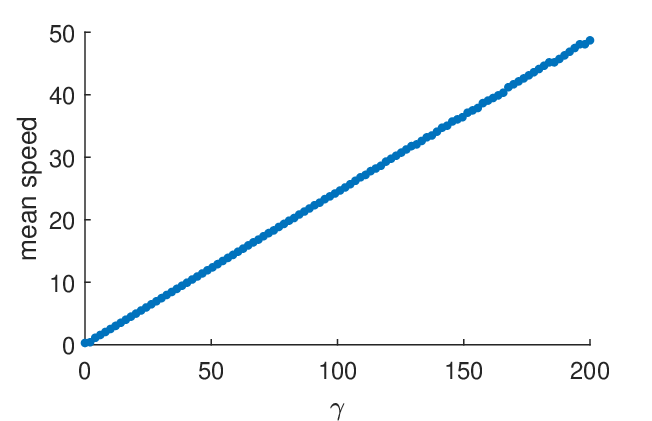}
    \caption{\justifying{The mean speed of a trajectory converging to a fixed point has been plotted against $\gamma$. The mean speed increases linearly with $\gamma$ with a 0.25 slope.}} 
    \label{fig4}
\end{figure}

We divide this distance by the time step to numerically calculate the speed. The initial state is taken very close to $|1\rangle$ state, and Hamiltonian $H = \sigma_z -\frac{\mathbb{I} - \sigma_z}{2}$ is applied to evolve it using \eqref{eq:07} so that the state finally goes to $|0\rangle$. Note that this Hamiltonian is the same as $H_{s_{-}}(t)$ except for the switch function $f(t)$. The mean speed has been shown for different values of $\gamma$ in FIG.~\ref{fig4}. It means that $\gamma$ dictates how fast the states converge to the fixed points, just like the parameter $\Gamma_k$ in the Lindblad master equation \eqref{eq:04}. Thus, $\gamma$ can be interpreted as the strength of coupling between system and apparatus.

Hence, in this section, we have shown that with an appropriate choice of $H_{m_{+}}$ or $H_{m_{-}}$, any state on the Bloch sphere can be sent to one of the eigenvectors of $H_h$. Depending on where we add the complex number $-i\gamma$ in the diagonal, the state converges to the other eigenvector. But this is not really consistent with the final state $|\Psi'\rangle$ expression that was obtained in the von Neumann scheme \eqref{eq:03}. If $|0\rangle$ is coupled with one ancilla state (say, $|-\rangle_y)$, then its orthogonal state $\ket{1}$ must be coupled to $|+\rangle_y$ ancilla state. In the next section, we will check if we can obtain a similar result by expanding the dimensionality of the ancilla Hilbert space and, hence, of the full space.





\section{The complete embedding}\label{sec:5}
We saw that by embedding a two-dimensional non-Hermitian Hamiltonian into a four-dimensional Hermitian Hamiltonian (thus, the dimensionality of ancilla space is 2), we obtained a final state that was inconsistent with the von Neumann measurement result. Let us first see why this is so by writing the final state $|\Psi(t_f)\rangle$ in full, and then how expanding the ancilla Hilbert space can help us get close to our desired result.

\subsection{The need to expand the full Hilbert space}
Let us consider the Hilbert spaces of both the system and ancilla to be two-dimensional (as done previously), and so non-Hermitian $H_s(t)$ is $2\cross2$. From \eqref{eq:14}, projection and post-selection over $|-\rangle_y$ ancilla state yields the system state $|\psi(t)\rangle = \mathcal{T}e^{-i\int H_s(t)dt}|\psi(0)\rangle$. Say $H_s(t)$ drives $|\psi(t)\rangle$ to $|0\rangle$, then the final full state will look like 
$$|\Psi(t_f)\rangle = |0\rangle\otimes|-\rangle_y+\eta(t_f)|0\rangle\otimes|+\rangle_y$$

As mentioned previously, in order to replicate the von Neumann scheme of measurement, we must obtain orthogonal eigenvectors coupled to the two ancilla states. If we want to access both the fixed points in this dynamics ($|0\rangle$ and $|1\rangle$) simultaneously, then $\eta(t)|\psi(t)\rangle$ which is coupled to the $|+\rangle_y$ ancilla state should be orthogonal to $|\psi(t)\rangle$ and stay orthogonal throughout the evolution. This basically gives us two conditions
$$\langle\psi(t)|\eta(t)\psi(t)\rangle = 0$$
and 
$$\frac{d}{dt}(\langle\psi(t)|\eta(t)\psi(t)\rangle) = 0$$
Expanding the last equation and substituting from \eqref{eq:1}, we get
$$i\frac{d}{dt}\eta(t) = H_s^{\dagger}(t)\eta(t) - \eta(t)H_s(t)$$
which is similar to the evolution equation for $M(t)$ \eqref{eq:9}. And so the evolution for $\eta(t)$ is also 
$$\eta(t) = e^{-i\int H_s^{\dagger}(t)dt}\eta(0) e^{i\int H_s(t)dt}$$
Let $M(0) = m_0\mathbb{I} \Rightarrow \eta(0) = \sqrt{m_0-1}\mathbb{I}$  as before. If we substitute the above time evolution expression in the definition of $M(t) = \eta^{\dagger}(t)\eta(t) + \mathbb{I}$, we get
$$M(t) = [(m_0-1)e^{2i\int (H_s(t) - H_s^{\dagger}(t))dt}+1]\mathbb{I}$$
which does not match the expression for $M(t) = m_0\mathbb{I} e^{i\int (H_s(t) - H_s^{\dagger}(t))dt}$. Thus, this is a contradiction, and so we cannot achieve simultaneous orthogonal evolution in the system subspace by dilation of a 2-dimensional non-Hermitian Hamiltonian into a 4-dimensional Hermitian one.


\subsection{Eight-dimensional embedding}
Let us increase the Hilbert space dimensionality of the ancilla, 
(hence of the full Hamiltonian $H_{sa}(t)$) and see if we can find an embedding to achieve orthogonal subspace dynamics. We consider a 4-level ancilla that spans a 4-dimensional Hilbert space. We name these four orthonormal ancilla states as $|0\rangle$, $|1\rangle$, $|2\rangle$, and $|3\rangle$. The total state $|\Psi(t)\rangle$ is an 8-dimensional state written as
$$|\Psi(t)\rangle = |\psi(t)\rangle \ket{0}  + \eta(t)\ket{\psi(t)}\ket{1} + |\chi(t)\rangle\ket{2}+ \zeta(t)\ket{\chi(t)}\ket{3} $$
where $\zeta(t)$ is a time-dependent operator like $\eta(t)$ and $\ket{\chi(t)}$ is some state in the 2-dimensional Hilbert space. Just as $\eta(t)|\psi(t)\rangle$ acts as sort of a ``dump" for non-Hermitian dynamics to come into the picture for state $\ket{\psi(t)}$, likewise $\zeta(t)\ket{\chi(t)}$ would be a ``dump" state to enable orthogonal non-Hermitian evolution for $\ket{\chi(t)}$. 

The state $\ket{\psi(t)}$ follows Eq.~\eqref{eq:1} with $H_s(t) = H_{s_{-}}(t)$ being the two-dimensional non-Hermitian Hamiltonian which drives $\ket{\psi(t)}$ to one of it's eigenvectors. Then, we would demand that 
\begin{equation}\label{eq:25}
    i\frac{d}{dt}|\chi(t)\rangle = H_{s_{+}}(t)|\chi(t)\rangle
\end{equation}
so that the state $\ket{\chi(t)}$ converges to a final state which is orthogonal to where $\ket{\psi(t)}$ goes to. We can expand the 8-dimensional $H_{sa}(t)$ as 
\begin{align}
    \label{eq:26}
    \begin{split}
         H_{sa}(t) = H_{sa}^{(00)}(t)\otimes\ket{0}\bra{0} + H_{sa}^{(01)}(t)\otimes\ket{0}\bra{1}+\dots\\
   +H_{sa}^{(33)}(t)\otimes\ket{3}\bra{3} = \sum_{i,j=0}^{i,j=3} H_{sa}^{(ij)}(t)\otimes\ket{i}\bra{j}
    \end{split}
\end{align} 
And the Hermiticity condition dictates that
$$(H_{sa}^{(ij)})^{\dagger}(t) = H_{sa}^{(ji)}(t)$$
for $i,j=0, 1, 2, 3$. Substituting \eqref{eq:1}, \eqref{eq:25}, \eqref{eq:26} in \eqref{eq:3} and then equating the operators, we will get very similar conditions on the Hamiltonian blocks as \eqref{eq:12} and \eqref{eq:13} and some more. Also, we get exactly the same evolution for $\eta(t)$ as in \eqref{eq:7} (except that $H_s$ is replaced by $H_{s_{-}}$), and thus, we can define $M(t)$ in the same way as before. The evolution for $\zeta(t)$ comes out as
\begin{equation}\label{eq:27}
    i\frac{d}{dt}[\zeta^{\dagger}(t)\zeta(t)] = H_{s_{+}}^{\dagger}(t)[\zeta^{\dagger}(t)\zeta(t)+\mathbb{I}]-[\zeta^{\dagger}(t)\zeta(t)+\mathbb{I}]H_{s_{+}}(t)
\end{equation}
which is very similar to $\eta(t)$ evolution except that $H_s{_{+}}$ comes in place of $H_{s_{-}}$. We define $N(t) = \zeta^{\dagger}(t)\zeta(t) +\mathbb{I}$ so that
$$i\frac{d}{dt}N(t) = H_{s_{+}}^{\dagger}(t)N(t)-N(t)H_{s_{+}}(t)$$
\begin{equation}\label{eq:28}
    \Rightarrow N(t) = \mathcal{T}e^{-i\int H_{s_{+}}^{\dagger}(t)dt}N(0) \Bar{\mathcal{T}}e^{i\int H_{s_{+}}(t)dt}
\end{equation}
and $\zeta(t) = V(t)(N(t) - \mathbb{I})^{\frac{1}{2}}$ where $V(t)$ is some unitary operator. We will drop the time dependencies in all the matrices from here on, as it is implied. 

Like before, we make very similar choices for simplification: $U = V = \mathbb{I}$ such that $\eta$ and $\zeta$ are both Hermitian and commute with $M$ and $N$, respectively. Also, we take 
\begin{equation}\label{eq:29}
    H_{sa}^{(11)} = \{H_{s_{-}}+[i\frac{d}{dt}\eta+\eta H_{s_{-}}]\eta\}M^{-1}
\end{equation}
which is same as \eqref{eq:10}. Also,
\begin{equation}\label{eq:30}
    H_{sa}^{(33)} = \{H_{s_{+}}+[i\frac{d}{dt}\zeta+\zeta H_{s_{+}}]\zeta\}N^{-1}
\end{equation}
which is analogous to the previous choice. Previously, it was noted that such a choice yielded the diagonal blocks of $H_{sa}$ to be the same and the off-diagonal blocks to be negatives of each other. Very similarly, we obtain here
\begin{equation}\label{eq:31}
    H_{sa}^{(00)} = H_{sa}^{(11)} \hspace{1cm} H_{sa}^{(22)} = H_{sa}^{(33)}
\end{equation}
Ancilla states $|0\rangle$ and $|1\rangle$ contained the expression for $\psi$, and we see that the corresponding blocks come out to be the same in $H_{sa}$. Similarly, blocks 22 and 33 that represent $\chi$ come out to be the same. Also, the corresponding off-diagonal blocks come out to be negatives of each other, and the expressions are similar to \eqref{eq:13}. 
\begin{equation}\label{eq:32}
    H_{sa}^{(01)} = [H_{s_{-}}\eta-\eta H_{s_{-}}-i\frac{d}{dt}\eta]M^{-1} =-H_{sa}^{(10)} 
\end{equation}
\begin{equation}\label{eq:33}
    H_{sa}^{(23)} = [H_{s_{+}}\zeta-\zeta H_{s_{+}}-i\frac{d}{dt}\zeta]N^{-1} =-H_{sa}^{(32)} 
\end{equation}
$H_{sa}^{(13)}$ also comes out to be of the exact form as above. 
\begin{equation}\label{eq:34}
    H_{sa}^{(13)} = H_{sa}^{(23)} = - H_{sa}^{(31)}
\end{equation}
The other blocks look like:
\begin{equation}\label{eq:35}
    H_{sa}^{(12)} = H_{sa}^{(33)}-H_{s_{+}} = (H_{sa}^{(21)})^{\dagger}
\end{equation}
\begin{equation}\label{eq:36}
    H_{sa}^{(02)} = \eta(H_{s_{+}}-H_{sa}^{(33)}) = (H_{sa}^{(20)})^{\dagger}
\end{equation}
\begin{equation}\label{eq:37}
    H_{sa}^{(30)} = H_{sa}^{(23)}\eta=(H_{sa}^{(03)})^{\dagger}
\end{equation}
\begin{figure}
    \centering
    \includegraphics[scale=1]{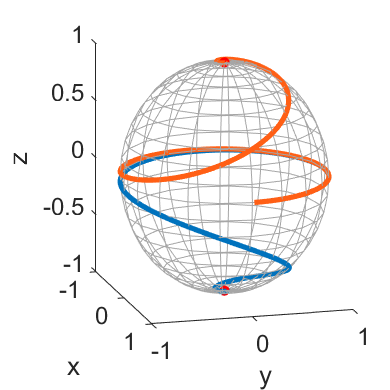}
    \caption{\justifying{Bloch sphere trajectories of $\ket{\psi(t)}$ and $\ket{\chi(t)}$ by taking projection and post-selecting over $\ket{0}$ and $\ket{2}$ states of ancilla. $|\psi\rangle$ and $|\chi\rangle$ states simultaneously converge to the North and South pole, respectively. The non-Hermitian Hamiltonians used in the embedding are $H_{s_{-}}(t)$ (driving $\psi$) and $H_{s_{+}}(t)$ (driving $\chi$) mentioned in \eqref{eq:21}. $|\psi(0)\rangle = |\chi(0)\rangle = |+\rangle$, $\gamma=1.5$ and $m_0=1.0005$.}}
    \label{fig:5}
\end{figure}
Thus we now have the expression for the full $8\cross8$ Hermitian $H_{sa}$ in terms of $H_{s_{-}},H_{s_{+}}, \eta, \zeta, M$ and $N$. Again, we have the constraint that the eigenvalues of $M$ and $N$ must be greater than one throughout the evolution. 

Projective measurement on $\Psi(t)$ and then post-selecting over $\ket{0}$ and $\ket{1}$ states of the ancilla will give us 
$$|\psi(t)\rangle= \mathcal{T}e^{-i\int H_{s_{-}}(t) dt}\ket{\psi(0)}$$
$$|\chi(t)\rangle= \mathcal{T}e^{-i\int H_{s_{+}}(t) dt}\ket{\chi(0)}$$
so that $\ket{\psi(t)}$ and $\ket{\chi(t)}$ will finally end up orthogonal to each other.

FIG.~\ref{fig:5} shows the trajectories of states $\ket{\psi}$ and $\ket{\chi}$ as they evolve on the Bloch sphere. The $H_{s_{\pm}}$ has been taken same as in equation \eqref{eq:21} and $H_{sa}$ has been constructed using equations \eqref{eq:29}-\eqref{eq:37}. Both $M(0)=N(0) = m_0\mathbb{I}$ where $m_0=1+\epsilon$ such that $\epsilon$ is very small. And we know from our analysis in Sec.~\ref{sec:3} that the lower eigenvalue of $M$ and $N$ will remain constant when evolved via $H_{s_{\pm}}$. The ancilla states have been taken as tensor products of eigenstates of $\sigma_y$. So, $\ket{0}\equiv|-\rangle_y|-\rangle_y$, $\ket{1}\equiv|+\rangle_y|-\rangle_y$, $\ket{2}\equiv|-\rangle_y|+\rangle_y$,  and $\ket{3}\equiv|+\rangle_y|+\rangle_y$.

This shows that in order to obtain simultaneous evolution towards orthogonal fixed points (or any two fixed points for that matter) in the system subspace, one would need to embed the 2-dimensional non-Hermitian Hamiltonians into at least an 8-dimensional Hermitian Hamiltonian. Let us now see how we can generalize our protocol for $N$-dimensional system Hilbert space.

\subsection{$N$-dimensional embedding}
Let us consider an $N$-level system state. From the previous subsection, it is clear that in order to achieve the orthogonal convergence to each of the eigenstates, one would need the target state and a `dump' state for each of the eigenstates. In that case, the ancilla Hilbert space must have a dimensionality of $2N$. So, the dimension of the full system-ancilla Hilbert space is $2N^2$.

The non-Hermitian Hamiltonians driving the initial state to orthogonal eigenvectors are denoted by $H_i$ where $i$ runs from 1 to $N$. The diagonal Hamiltonian $H_i$ is such that its $i^{\text{th}}$ diagonal entry has no imaginary part while all other diagonal entries have negative imaginary parts. For simplicity, we can construct $H_i$ as 
\begin{equation}
    H_i = \mathbb{I}_N + i\gamma f(t) (P_i - \mathbb{I}_N)
\end{equation}
where $f(t)$ is same as Eq.~\ref{eq:ft} and $P_i =|e_i\rangle\langle e_i|$ is the projection operator to $i^{\text{th}}$ eigenvector, i.e., all other entries 0 except the $i^{\text{th}}$ diagonal entry which is 1. The ancilla states are $|1\rangle, |2\rangle, \dots,|2N\rangle$ such that the total state $|\Psi\rangle$ looks like 
$$|\Psi\rangle = |\psi_1\rangle|1\rangle+ \eta_1|\psi_1\rangle|2\rangle+ |\psi_2\rangle|3\rangle+ \eta_2|\psi_2\rangle|4\rangle+\dots+|\psi_N\rangle|2N-1\rangle+ \eta_N|\psi_N\rangle|2N\rangle$$
$$\Rightarrow |\Psi\rangle = \sum_{i=1}^N (|\psi_i\rangle|2i-1\rangle + \eta_i|\psi_i\rangle|2i\rangle)$$
such that the even ancilla indices will entangle with the dump states at the end of the measurement, and the odd ancilla index positions will correspond with the eigenvectors on the system side. We also have $|\psi_1(0)\rangle = |\psi_2(0)\rangle=\dots = |\psi_N(0)\rangle = c_1|e_1\rangle+c_2|e_2\rangle+\dots + c_N|e_N\rangle$. As before, we demand that 
$$i\frac{d}{dt}|\psi_i\rangle = H_i|\psi_i\rangle$$
And the full state is evolved by the full Hamiltonian as in Eq.~\ref{eq:3}. Performing similar calculations as before, we obtain
$$ i\frac{d}{dt}M_i = H_i^{\dagger}M_i-M_iH_i$$
where $M_i = \eta_i^{\dagger}\eta_i+\mathbb{I}_N$, which are again similar to the equations we already obtained previously. We can write $H_{sa}$ as 
$$H_{sa} = \sum_{i,j=1}^{i,j=2N} H_{sa}^{(i,j)}\otimes\ket{i}\bra{j}$$
Solving for the full Hamiltonian $H_{sa}$, we will obtain similar expressions to the ones obtained in the previous subsection for the `tridiagonal blocks' (Eqs.~\eqref{eq:29}-\eqref{eq:33}). These tridiagonal blocks of $H_{sa}$ would follow the relations 
\begin{equation}\label{eq:46}
    H_{sa}^{(2i-1,2i-1)} = H_{sa}^{(2i,2i)}\\
    H_{sa}^{(2i-1,2i)} = -H_{sa}^{(2i,2i-1)}
\end{equation}
where each of the $H_{sa}^{(i,j)}$ are $N\cross N$ blocks. The other entries (like Eqs.~\eqref{eq:34}-\eqref{eq:37}) can be obtained by the same procedure. These will be expressed in terms of $\eta_i$, $H_i$, and the tridiagonal matrices mentioned in Eq.~\eqref{eq:46}.

Operating this Hamiltonian $H_{sa}$ on $|\Psi(t)\rangle$ will give us a final state where we have convergence to each of the $N$ eigenvectors. Applying the projection operator like in Eq.~\eqref{eq:proj}, we can see how each of the states $|\psi_i\rangle$ finally converges to $|e_i\rangle$ eigenvector. 

Let us discuss in some detail about the final state $|\Psi(t_f)\rangle$ that is obtained after operating $H_{sa}$ on $|\Psi(t)\rangle$. This has been done in the next section, continuing with the 8-dimensional embedding example.

\section{Discussion}\label{sec:6}
If one expands the full state $|\Psi(t)\rangle$ (obtained in subsection 5.2) at time $t_f$ (i.e., after subspace convergence to orthogonal eigenvectors), one will get
\begin{equation}\label{eq:42}
    |\Psi(t_f)\rangle = c'_1|0\rangle |0\rangle  + \eta(t_f)|0\rangle|1\rangle +  c'_2|1\rangle|2\rangle+\zeta(t_f)|1\rangle|3\rangle.
\end{equation}
Since $\eta(t_f)|0\rangle$ and $\zeta(t_f)|1\rangle$ are non-zero, one cannot obtain the entangled final state as proposed by von Neumann \eqref{eq:03}. We observe that the ratio of the probability coefficients remains the same. If the initial state is $|\psi\rangle = c_1|0\rangle+c_2|1\rangle$, then 
$$\dfrac{c'_1}{c'_2} = \dfrac{c_1}{c_2}.$$
We also notice that as $m_0$ gets closer to 0, $c'_{1,2} \approx \frac{1}{2}c_{1,2}$ and as $m_0$ gets larger than 0, $c'_{1,2}$ get farther from $c_{1,2}$.

Thus, in order to obtain a state that looks like Eq.~\ref{eq:03}, one can do a normalized projection over ancilla states $|0\rangle$ and $|2\rangle$ in Eq.~\eqref{eq:42}. This projection operator looks like 
$$\mathcal{P'} = \dfrac{\mathbb{I}\otimes|0\rangle \langle 0| + \mathbb{I}\otimes|2\rangle \langle 2|}{\sqrt{\langle\Psi(t)|\mathbb{I}\otimes|0\rangle\langle 0| + \mathbb{I}\otimes|2\rangle \langle 2|\Psi(t)\rangle}}$$
Applying this operator on $|\Psi(t_f)\rangle$, we get
$$\mathcal{P'}|\Psi(t_f)\rangle=c_1|0\rangle|0\rangle+c_2|1\rangle|2\rangle$$
This is the total state that is exactly the same as proposed by von Neumann at the end of the unitary operation $\mathcal{U}$. In our case, this state is formed after operating $e^{-i\int H_{sa} dt}$ on $|\Psi(0)\rangle$ (which is unitary) and then the non-unitary projection $\mathcal{P'}$. 

\begin{figure}
    \centering
    \includegraphics[scale=1]{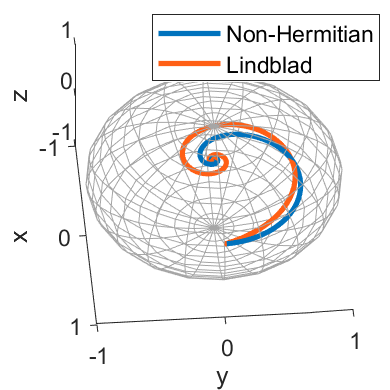}
    \caption{\justifying{Bloch sphere trajectory for non-Hermitian dynamics (blue) and the Lindblad dynamics (orange) starting from the same initial state. The steady state in both cases is the same.}}
    \label{fig:6}
\end{figure}

In Fig.~\ref{fig:6}, we have compared the Bloch sphere trajectories obtained from Lindblad master equation \eqref{eq:04} and non-Hermitian dynamics as discussed in the previous section. It is important to note that for non-Hermitian dynamics, we have first applied the projection operator $\mathcal{P'}$ on $\Psi(t_f)$ and then taken the partial trace to get the system density matrix $\rho_s(t_f)$. The initial state has been taken as $|\psi(0)\rangle = \sqrt{\dfrac{2}{3}}|0\rangle+\dfrac{1}{\sqrt{3}}|1\rangle$. For the Lindblad dynamics, $\Gamma=0.3$; $H_s$ and the ladder operator $L$ both have been taken as $\sigma_z$. This means that we are measuring the spin along the $z$ direction. For non-Hermitian dynamics, we have used the same operators and the same values as in Fig.~\ref{fig:5}.

It can be seen that an initially pure state becomes a mixed state (decoherence) in both cases. The steady state in both cases is the same. Thus, effectively, using the embedding protocol and a particular projection operator, we have been able to emulate the Lindblad dynamics via non-Hermitian Hamiltonian dynamics.

It can be argued that we should have used $H_s(t)$ and $H_s^{\dagger}(t)$ to simulate exact orthogonal dynamics in the last section. If that were possible, we would have obtained forward and backward propagation of a state in the same system. But using $H_s(t)$ and $H_s^{\dagger}(t)$ would not respect the eigenvalue constraint of both $M(t)$ and $N(t)$ simultaneously. If $H_s(t)$ is such that the lowest eigenvalue of $M(t)$ remains above one throughout the evolution, $H_s^{\dagger}(t)$ would not be able to do that for $N(t)$. At least, this is true for the choice of $H_s(t)$ that we have made. Some different choices of Hamiltonians or initial choices of $M$ and $N$ can be looked into to simulate this dynamics.  

Using a similar procedure of deriving the blocks of the $H_{sa}(t)$ matrix as shown in the last section, it can be deduced that a 6-dimensional Hilbert space for the total system-ancilla state would not be sufficient for achieving simultaneous orthogonal fixed points. We would have to go to 8 dimensions.

\section{Conclusion and Outlook}\label{sec:7}
In this paper, we have tried to emulate the measurement scheme as proposed by von Neumann --- that the total state goes from a product to an entangled state via a unitary operation. The Lindblad master equation brings about decoherence in the system subspace and mimics the measurement scheme when the Ladder operators are Hermitian. We tried to find if there is a Hamiltonian formalism to obtain similar results. 

We have shown that the dilation of a specific non-Hermitian Hamiltonian $H_s$ with complex eigenvalues into a higher dimensional Hermitian Hamiltonian $H_{sa}$ can be used to access the fixed points (eigenvectors) of $H_s$. We showed that in order for this embedding protocol to work, the eigenvalues of $H_s$ must have non-positive imaginary parts. The diagonal nature of the non-Hermitian Hamiltonian enables us to extend the protocol for any N-level system state. Using the example of a two-dimensional Hamiltonian embedded into a four-dimensional one, we showed that we obtain an entangled state at the end, and the speed of convergence to the fixed point can be controlled by a single parameter $\gamma$. We also proved that the dilation to four-dimensional Hilbert space would not result in accessing both the fixed points of the two-dimensional non-Hermitian Hamiltonian simultaneously. In order to do that, we need to dilate it to an eight-dimensional Hermitian Hamiltonian. As a result, we get two extra ``dump" states in our total final state. Because of the non-zero dump states, we could not achieve exactly the same expression of the final state as proposed by von Neumann. We applied a particular projection operator to get the desired result. We also showed how this procedure could be generalized to an $N$-dimensional system Hilbert space. Finally, we compared the dynamics obtained from our embedding scheme with the evolution from the Lindblad master equation and found that we obtain the same steady state in both cases. 


The proposed model partially emulates the measurement results obtained from the von Neumann scheme and the Lindblad master equation. If we could obtain the final entangled state after measurement \eqref{eq:02} without the projection operator $\mathcal{P'}$, then our embedding protocol is a valid emulation of the Lindblad master equation used in the context of measurement. There is a lot of freedom with regard to $M$ and $N$ matrices, which we have not employed. This can be a future line of work.\bigskip\\
\textbf{Data Availability Statement}\medskip\\
No new data were created or analyzed in this study.  

\bigskip
\ack
We would like to thank Sourin Das, Rangeet Bhattacharyya, Anant Varma, Arnab Acharya, and Akhil Bhartiya for useful discussions. S.B. acknowledges the J.C. Bose National Fellowship provided by SERB, Government of India, Grant No. JBR/2020/000049. G.S. thanks the Department of Science and Technology (DST), Government of India, for support through an INSPIRE Fellowship.\bigskip\\
\textbf{ORCID iDs}\medskip\\
Gurpahul Singh \href{https://orcid.org/0009-0005-5900-2021}{https://orcid.org/0009-0005-5900-2021}\\
Ritesh K. Singh \href{https://orcid.org/0000-0001-7838-6191}{https://orcid.org/0000-0001-7838-6191}\\
Soumitro Banerjee \href{https://orcid.org/0000-0003-3576-0846}{https://orcid.org/0000-0003-3576-0846}
\medskip


\nocite{*}

\section*{References}
\bibliographystyle{unsrt}
\bibliography{main}

\end{document}